\documentclass[12pt,preprint]{aastex}
\slugcomment{KSUPT -- 08/2  \quad  March 2008}

\begin{document}

\title{Constraints on Dark Energy from Galaxy Cluster Gas Mass Fraction versus Redshift data}

\author{Lado Samushia\altaffilmark{1,2,3} and Bharat Ratra\altaffilmark{1,4}}

\altaffiltext{1}{Department of Physics, Kansas State University, 116 Cardwell Hall, Manhattan, KS 66506}
\altaffiltext{2}{National Abastumani Astrophysical Observatory, 2A Kazbegi Ave, GE-0160 Tbilisi, Georgia}
\altaffiltext{3}{email: lado@phys.ksu.edu}
\altaffiltext{4}{email: ratra@phys.ksu.edu}

\begin{abstract}
We use the \citet{all08} galaxy cluster gas mass fraction versus redshift data  to constrain parameters of three different dark energy models: a cosmological constant dominated one ($\Lambda$CDM); the XCDM parameterization of dark energy; and a slowly-rolling scalar field model with inverse-power-law potential energy density. (Instead of using the Monte Carlo Markov Chain method, when integrating over nuisance parameters we use an alternative method of introducing an auxiliary random variable.)  The resulting constraints are consistent with, and typically more constraining than, those derived from other cosmological data. A time-independent cosmological constant is a good fit to the galaxy cluster data, but slowly evolving dark energy cannot yet be ruled out.
\end{abstract}


\section{Introduction}

Observations over the last decade have established that the cosmological expansion is accelerating. In the context of general relativity this requires that the cosmological energy budget be dominated by dark energy \citep[for reviews see, e.g., ][]{cop07, pad06, ish07, uza07, lin07, rat08}. This hypothetical construct --- dark energy --- is an enigma. It is not yet clearly established whether dark energy is Einstein's cosmological constant $\Lambda$ \citep[e.g.,][]{pee84}, or whether it evolves slowly in time and varies weakly in space \citep[e.g.,][]{pee88, rat88}. While Type Ia supernova (SNIa) apparent magnitude measurements as a function of redshift indicate accelerated cosmological expansion \citep{rie98, per99}, SNIa data are as yet unable to unambiguously constrain dark energy  \citep[see, e.g., ][]{ala07, nes07, sha07, zhan07, wu08, ish08}. Future SNIa data will improve the constraints \citep[e.g.,][]{pod01a} and could resolve some of the current differences in results from different SNIa data sets. 

The results of the SNIa test are confirmed by a test based on cosmic microwave background (CMB) anisotropy data that must assume the cold dark matter (CDM) model for structure formation (see Peebles \& Ratra 2003 and references therein for a discussion of apparent problems with the CDM model). CMB anisotropy data is consistent with the Universe having negligible spatial curvature \citep[see, e.g.,][]{pod01b, dur03, muk03, pag03, spe07, dor07a}, under the assumption that dark energy does not evolve in time \citep[e.g.,][]{wri06, teg06, zhao07, ich07, wang07}. In combination with low measured non-relativistic matter density (Chen \& Ratra 2003b and references therein), negligible spatial hypersurface curvature indicates the presence of dark energy.

There are many different models of dark energy.\footnote{In this paper we assume general relativity is an adequate description of gravitation on cosmological scales. For discussions of accelerated cosmological expansion in modified gravity models see, e.g., \citet{mov07}, \citet{fay07}, \citet{ame07}, \citet{wangetal07}, \citet{wan07}, \citet{wang07a}, and \citet{dem07}. We also assume that dark energy and dark matter are uncoupled. For discussions of coupled or unified dark energy and dark matter models see, e.g., \citet{bon06}, \citet{wu07}, \citet{guo07}, \citet{wei07}, and \citet{oli07}. For other models see, e.g., \citet{bar06}, \citet{bra07}, \citet{dut07}, \citet{wuetal07}, \citet{gra07}, and \citet{neu07}.} In this paper we consider three: standard $\Lambda$CDM, the XCDM parameterization, and a slowly rolling scalar field dominated one ($\phi$CDM). In the $\Lambda$CDM model the late-time Universe is dominated by a cosmological constant $\Lambda$ with time-independent energy density $\rho_\Lambda$ \citep{pee84}. In $\phi$CDM the dark energy is a slowly rolling scalar field $\phi$; in the model we consider the scalar field potential energy density $V(\phi)\propto \phi^{-\alpha}$, where $\alpha$ is a nonnegative parameter \citep{pee88}. We also consider the XCDM parameterization of the dark energy equation of state; here dark energy is modeled as a fluid with an equation of state which relates the fluid pressure $p=\omega_{\rm x}\rho$ to its energy density $\rho$ where $\omega_{\rm x}$ is a time-independent negative parameter. This approximation is inaccurate in the scalar field dominated epoch \citep{rat91}. In all three models, other contributors to the Universe's current energy budget include CDM and baryonic matter. In the $\phi$CDM and XCDM cases spatial hypersurfaces are taken to be flat, while spatial curvature is treated as a cosmological parameter in the $\Lambda$CDM model we consider.

It is important to confirm and strengthen the SNIa and CMB test results by using additional techniques. This will allow for consistency checks as well as possibly identifying systematic effects in a particular data set. Other promising current tests include the angular size of radio sources and quasars as a function of redshift \citep[e.g.,][]{che03a, pod03, dal05, dal07}, strong gravitational lensing \citep[e.g.,][]{cha04, alc05, fed07, lee07, ogu07}, measurements of the Hubble parameter as a function of redshift \citep[e.g.,][]{sam06, sen08, laz07b, wei07, sam07}, and large-scale structure baryon acoustic oscillation measurements \citep[e.g.,][]{dor07b, par07, per07, lim07}. For reviews of the observational situation see \citet{kur07}, \citet{wang07b}, and \citet{laz07a}. Current data favor dark energy that does not evolve, but do not yet strongly rule out evolving dark energy \citep[see, e.g.,][]{rap05, wil06, dav07}.

In this paper we use new galaxy cluster gas mass fraction versus redshift data \citep[hereafter A08]{all08} to constrain parameters of the three dark energy models mentioned above. Earlier galaxy cluster gas mass fraction data \citep{all04} have been used to constrain these and other models \citep*[e.g.,][]{che04, rap05, alc05, can06, zha06, wei07}.

If rich galaxy clusters are large enough to have matter content that fairly samples that of the Universe then the baryonic to total mass ratio in clusters is equal to the cosmological baryonic mass fraction ratio of $\Omega_{\rm b}$ and $\Omega_{\rm m}$, the baryonic and nonrelativistic mass density parameters. The baryonic mass in clusters is dominated by the x-ray emitting gas which can be measured through x-ray observations. Combined with an estimate of $\Omega_{\rm b}$ from other observations, these measurements can be used to constrain $\Omega_{\rm m}$. The gas mass fraction in large relaxed clusters is expected to be independent of redshift and, since the observed x-ray temperature depends on the assumed distance to the cluster, this can be used to constrain cosmological parameters (\citealp{sas96, pen97}; for recent discussions of the test see, e.g., \citealp{ett06, nag07, cra07}).

In addition to depending on cosmological parameters, in a given model the predicted gas mass fraction depends on a number of ``nuisance" parameters that have to be marginalized over to derive the probability distribution function for the cosmological parameters of interest. We note that the likelihood function depends on certain functions of the ``nuisance" parameters and so introduce auxiliary random variables to describe these functions. This technique helps us to significantly reduce the computational time.

In next section we outline our computations. In Sec.\ 3 we present and discuss our results and conclude.

\section{Computation}

In our computations we follow A08. For a given cosmological model we compute predicted values of the gas mass fraction

\begin{equation}
f_{\rm gas}^{\rm th}(z, h, \Omega_{\rm m}, p, \Omega_{\rm b}, Q) = \frac{ K A \gamma b_0(1+\alpha_b z)} {1+s_0(1+\alpha_s z) } 
\left( \frac{\Omega_{\rm b}}{\Omega_{\rm m}} \right)
\left[ \frac{d_{\rm A}^{\Lambda}(z)}{d_{\rm A}(z)} \right]^{1.5},
\label{eq:fgas}
\end{equation}

\noindent
as a function of cluster redshift $z$, 4 cosmological parameters ($h, \Omega_{\rm m}, \Omega_{\rm b}$, and a parameter $p$, described below, that represents the dark energy model), and 7 parameters ($s_0$, $b_0$, $\alpha_s$, $\alpha_b$, $K$, $A$, $\gamma$) represented by $Q$ and related to modeling the cluster gas mass fraction.\footnote{These are discussed in depth by A08. Parameter $A$ is the angular correction factor and is close to unity for all redshifts and cosmologies of interest (A08) so we take $A=1$.} Here $h$ is the Hubble parameter in units of 100 $\rm km s^{-1} Mpc^{-1}$, $d_{\rm A}^{\Lambda}(z)$ is the angular diameter distance computed in the reference spatially-flat $\Lambda \rm{CDM}$ model (with $\Omega_{\rm m}=0.3$, $\Omega_\Lambda=0.7$, and $h=0.7$, where $\Omega_\Lambda$ is the cosmological constant density parameter), and $d_{\rm A}(z)$ is the angular diameter distance computed in the cosmological model of interest and depends on the model and $h$. Since the angular diameter distance $d_{\rm A}(z)$, and so $f_{\rm gas}^{\rm th}$, depends on the assumed dark energy model, we can compare predicted values of the gas mass fraction with measurements for clusters at redshift $z_i$ by constructing a $\chi^2=\sum_i(f_{\rm gas}^{\rm th}(z_i)-f_{\rm gas}^{\rm obs}(z_i))^2/\sigma_i^2$ function ($\sigma_i$ are the one standard deviation  measurement errors and the summation is over the 42 A08 clusters), and so constrain parameters of given dark energy models.

We construct a likelihood function $L=e^{-\chi^2/2}$, which depends on cosmological parameters like $\Omega_{\rm m}$ and those describing the dark energy model, as well as on the nuisance parameters. We marginalize over the nuisance parameters by multiplying the likelihood by the probability distribution function for the nuisance parameters and then integrating \citep[e.g.,][]{gan97}. The resulting probability distribution function depends only on two variables: $\Omega_m$ and a parameter $p$ describing the dark energy model. In the $\Lambda$CDM case $p$ is $\Omega_\Lambda$, in XCDM it is $\omega_{\rm x}$, and in $\phi$CDM it is $\alpha$. Since we consider only spatially-flat cosmologies for XCDM and $\phi$CDM models, two parameters $p$ and $\Omega_m$ completely describe the background evolution.

Since our initial likelihood function depends on 10 parameters in total (after setting $A=1$), to get to the two-dimensional probability distribution function we have to marginalize over 8 nuisance parameters ($h$, $\Omega_{\rm b}$, and the 6 parameters used to model the cluster gas mass fraction). To reduce computational time we use the following statistics results \citep[see, e.g.,][chap.~26]{ril02}. If two random variables $a$ and $b$ are independent and have probability distribution functions $P_a(x)$ and $P_b(x)$, then variables $c=ab$, $d=a/b$, and $f=F(a)$ are also random with probability distribution functions

\begin{equation}
P_c(x)=\int{\int{P_a(x')P_b(x'')\delta(x'x''-x) dx' dx''}}=\int{\frac{1}{|x'|} P_a(x') P_b(x/x')dx'},
\label{eq:mult}
\end{equation}

\begin{equation}
P_d(x)=\int{\int{P_a(x')P_b(x'')\delta(x'/x''-x) dx' dx''}}=\int{|x'| P_a(x x') P_b(x') dx'},
\label{eq:frac}
\end{equation}

\begin{equation}
P_f(x)=\left|\frac{dF^{-1}(x)}{dx}\right|P_a(F^{-1}(x)),
\end{equation}

\noindent
where $\delta(x)$ is a Dirac delta function. For example, since $d_{\rm A}$ in Eq.\ (\ref{eq:fgas}) is inversely proportional to $h$, we define an auxiliary variable $\Upsilon=K \gamma b_0 (\Omega_{\rm b}h^2)/\sqrt{h}$ and numerically compute a probability distribution function for it. This allows us to replace a five-dimensional integration (over $K$, $\gamma$, $b_0$, $\Omega_{\rm b}h^2$, and $h$) by a one-dimensional integral over a new variable $\Upsilon$ and so reduce computational time significantly.

Prior probability distribution functions for nuisance parameter are given in Table 1. The distribution functions for the 6 cluster gas mass fraction parameters are given A08, Table 4. Best fit values and confidence level constraints on dark energy parameters are sensitive to the assumed priors for the baryonic mass density and the Hubble parameter. Since different experiments give somewhat different estimates of these two, in our paper we use two prior sets to illustrate the differences. One set is from WMAP 3 year cumulative data \citep{spe07}, $h=0.73\pm0.03$ and $\Omega_{\rm b}h^2=0.0223\pm0.0008$, one standard deviation errors. The alternate set of priors we use is $h=0.68\pm0.04$ \citep{got01,che03c} and $\Omega_{\rm b}h^2=0.0205\pm0.0018$ \citep{fie06}, one standard deviation errors.

We define best fit values as pairs of values of cosmological parameters for which the likelihood function reaches its maximum. 1, 2, and 3 $\sigma$ confidence level contours are defined as sets of points in the parameter space with likelihood equal to $e^{-2.30/2}$, $e^{-6.17/2}$, and $e^{-11.8/2}$ of the maximum value of the likelihood respectively.

\section{Results and Discussion}
As a test we computed confidence level contours using our technique and the same prior set A08 used for $\Lambda$CDM and XCDM, Figs.\ 6 and\ 8 of A08. Our contours are in good agreement with those of A08.

Figures 1 to 3 show cluster gas mass fraction confidence level contours and best fit values for the three dark energy models and the two sets of priors for $\Omega_{\rm b}h^2$ and $h$ given in Table \ 1.  Compared to the constraints derived from the earlier \citet{all04} cluster gas mass fraction data \citep{che04}, the difference between the contours corresponding to the two prior sets (for $\Omega_{\rm b}h^2$ and $h$) is much reduced. The new constraints are almost as restrictive as the ones derived from SNIa data and more constraining than those derived from angular size versus redshift data or Hubble parameter versus redshift data.

Figure 1 shows constraints on $\Lambda$CDM. $\Omega_{\rm m}$ is better constrained than $\Omega_\Lambda$ and the results are in good qualitative accord with previous analyses. The best fit values are slightly away from a spatially-flat model.

Figure 2 shows constraints on the XCDM parameterization. Again, the energy density of nonrelativistic matter is fairly well constrained while the equation of state parameter is less constrained. The best fit values are again not exactly on the $\omega_{\rm x}=-1$ line, which corresponds to the spatially-flat $\Lambda$CDM case.

Figure 3 shows constraints on the $\phi$CDM model. $\Omega_{\rm m}$ is better constrained than $\alpha$. For both sets of priors there is an upper limit on $\alpha$. The best fit values are on the $\alpha=0$ line which corresponds to the spatially-flat $\Lambda$CDM case, but there is a big part of evolving dark energy parameter space that still is not ruled out.

\acknowledgements
We acknowledge support from DOE grant DE-FG03-99EP41093 and INTAS grant 061000017-9258.

\clearpage

\begin{deluxetable}{lccc}
\tablecaption{Prior probability distribution functions for nuisance parameters}
\tablehead{\colhead{}&\colhead{Parameter}&\colhead{Allowance}&\colhead{Distribution}}
\startdata
WMAP prior $\Omega_{\rm b}h^2$         & $\Omega_{\rm b}h^2$  & $0.0223\pm0.0008$             &Gaussian\\
WMAP prior $h$                         & $h$                  & $0.73\pm0.03$                 &Gaussian\\ 
Alternate prior $\Omega_{\rm b}h^2$    & $\Omega_{\rm b}h^2$  & $0.0205\pm0.0018$             &Gaussian\\
Alternate prior $h$                    & $h$                  & $0.68\pm0.04$                 &Gaussian\\ 
\enddata
\end{deluxetable}

\begin{figure}
\includegraphics[height=6.5in, width=7in, trim=0.3in 0in 0in 0in, clip, scale=1]{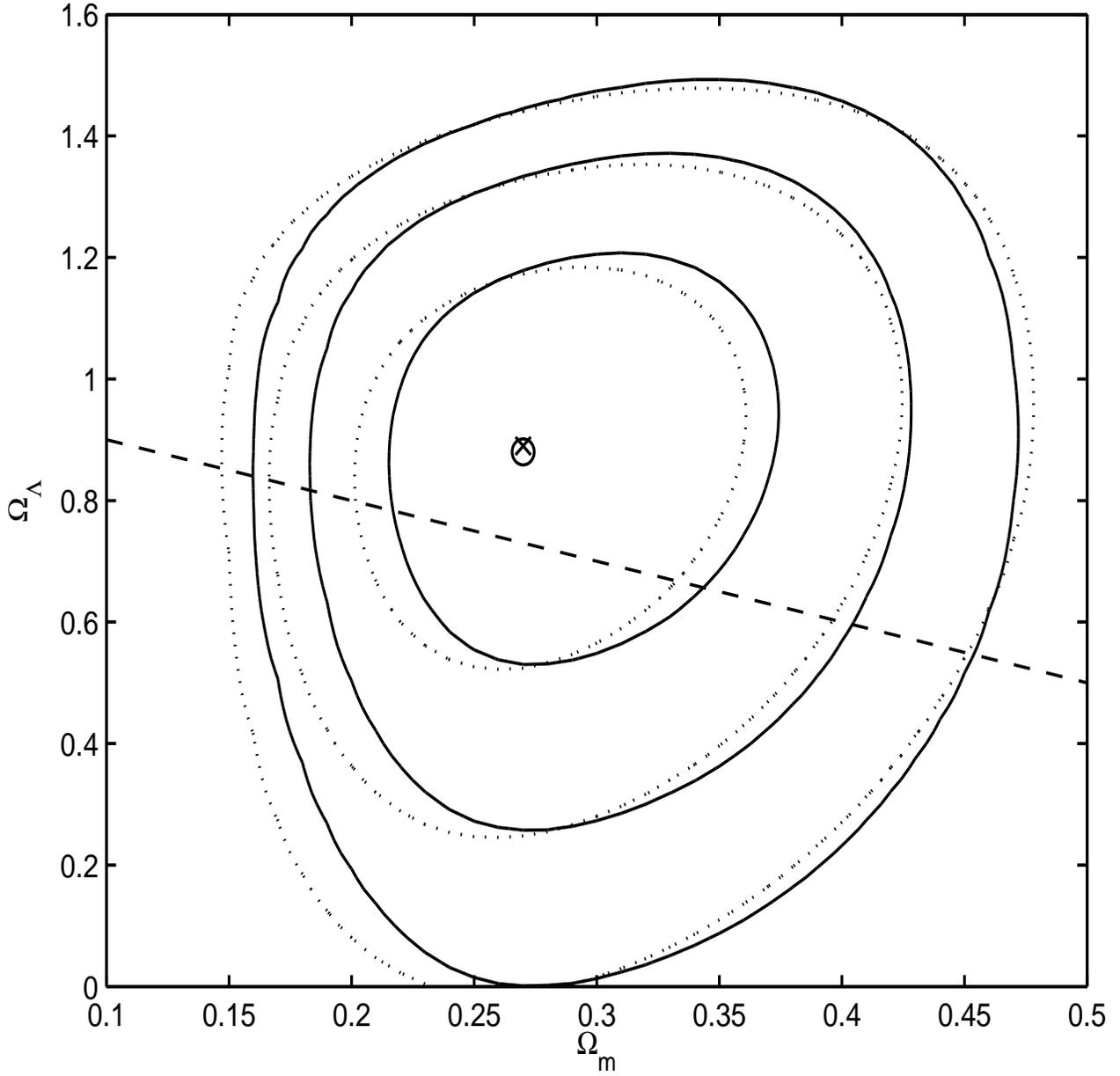}
\caption{1, 2, and 3$\sigma$ confidence level contours for the $\Lambda$CDM model. Solid lines ($\times$ for the best fit at $\Omega_{\rm m}=0.27$, $\Omega_\Lambda=0.89$) correspond to the WMAP prior and dotted lines ($\circ$ for the best fit at $\Omega_{\rm m}=0.27$, $\Omega_\Lambda=0.88$) correspond to the alternate prior. Dashed $\Omega_\Lambda=1-\Omega_{\rm m}$ line corresponds to the spatially-flat $\Lambda$CDM model.}
\end{figure}

\begin{figure}
\includegraphics[height=6.5in, width=7in,  trim=0.2in 0in 0in 0in, clip, scale=1]{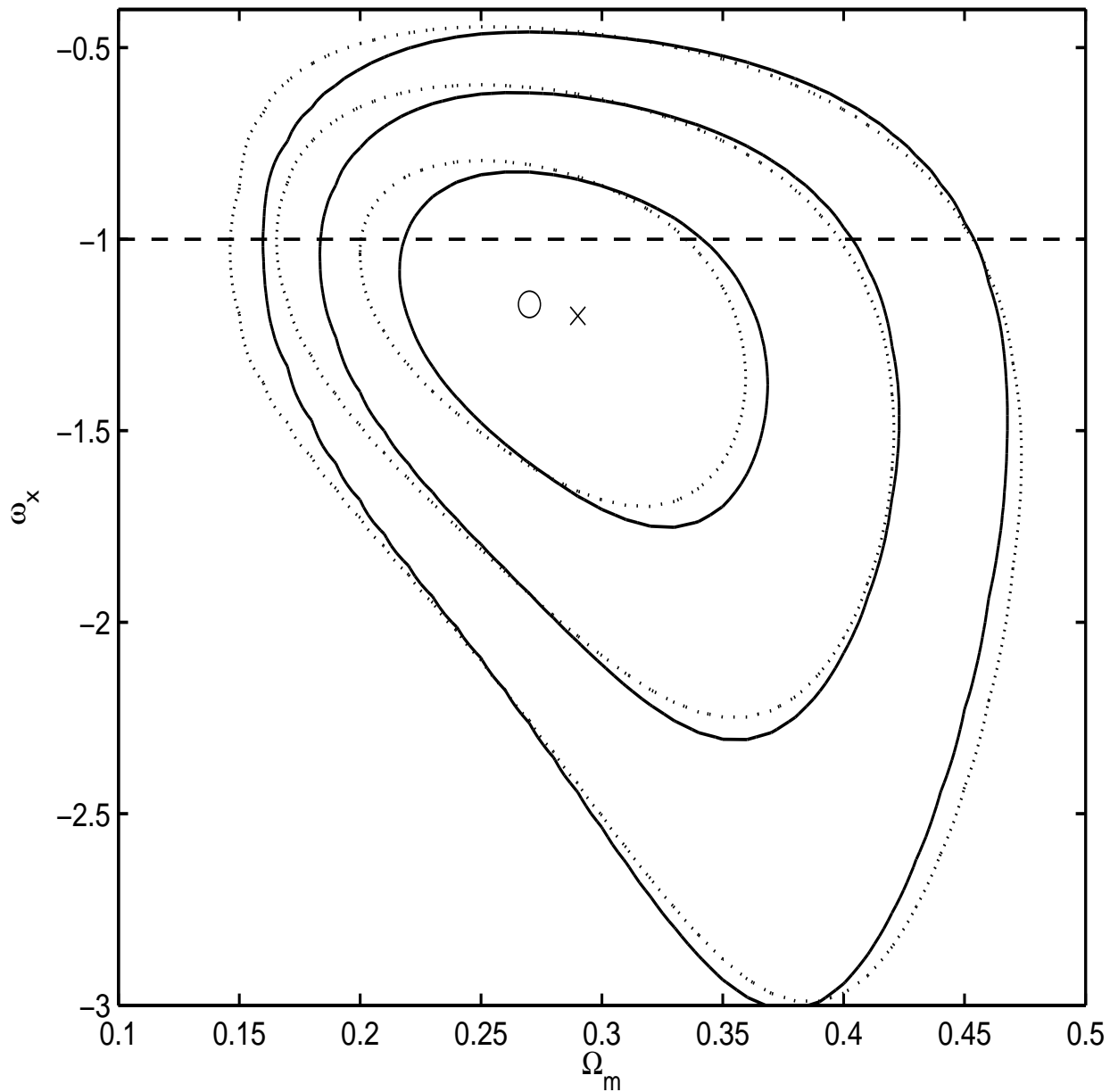}
\caption{1, 2, and 3$\sigma$ confidence level contours for the XCDM parameterization. Solid lines ($\times$ for the best fit at $\Omega_{\rm m}=0.29$, $\omega_{\rm x}=-1.2$) correspond to the WMAP prior and dotted lines ($\circ$ for the best fit at $\Omega_{\rm m}=0.27$, $\omega_{\rm x}=-1.17$) correspond to the alternate prior. Dashed $\omega_{\rm x}=-1$ line corresponds to the spatially-flat $\Lambda$CDM model.}
\end{figure}

\begin{figure}
\includegraphics[height=6.5in, width=7in,  trim=0.3in 0in 0in 0in, clip, scale=1]{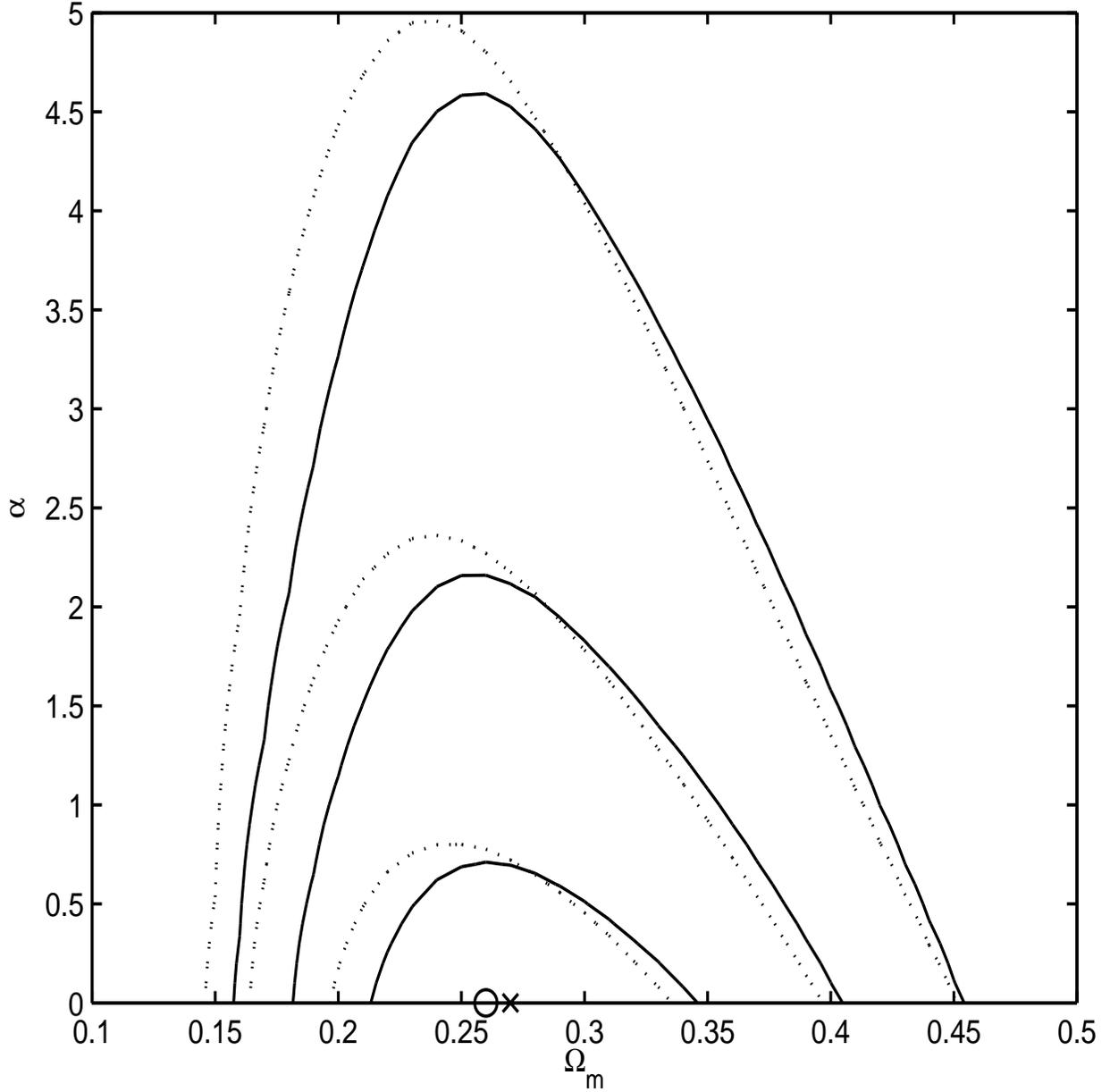}
\caption{1, 2, and 3$\sigma$ confidence level contours for the $\phi$CDM model. Solid lines ($\times$ for the best fit at $\Omega_{\rm m}=0.27$, $\alpha=0.0$) correspond to the WMAP prior and dotted lines ($\circ$ for the best fit at $\Omega_{\rm m}=0.26$, $\alpha=0.0$) correspond to the alternate prior. The $\alpha=0$ horizontal line corresponds to the spatially-flat $\Lambda$CDM model.}
\end{figure}

\end{document}